\documentstyle[12pt]{article}

\begin{document}

\begin{titlepage}
\vspace*{5mm}
\begin{center} {\Large \bf A new class of integrable diffusion--reaction
processes }\\ \vskip 1cm
{\bf M. Alimohammadi$^{a,b}$ \footnote {e-mail:alimohmd@theory.ipm.ac.ir},
N. Ahmadi$^a$}\\

\vskip 1cm
{\it $^a$ Physics Department, University of Tehran, North Karegar Ave.,} \\
{\it Tehran, Iran }\\
{\it $^b$ Institute for Studies in Theoretical Physics and Mathematics,}\\
{\it  P.O.Box 5531, Tehran 19395, Iran}
\end{center}
\vskip 2cm
\begin{abstract}
We consider a process in which there are two types of particles, $A$ and $B$,
on an infinite one--dimensional lattice. The particles hop to their adjacent
sites, like the totally asymmetric exclusion process (ASEP), and have also the
following interactions:
$A+B\rightarrow B+B$ and $B+A\rightarrow B+B$, all occur with equal rate. We
study this process by imposing four boundary conditions on ASEP master equation.
It is shown that this model is integrable, in the sense that its $N$--particle
S--matrix is factorized into a product of two--particle S--matrices and, more
importantly, the two--particle S--matrix satisfy quantum Yang--Baxter equation.
Using coordinate Bethe--ansatz, the $N$--particle
wavefunctions and the two--particle conditional probabilities are found exactly.

Further, by imposing four reasonable physical conditions on two--species
diffusion--reaction processes (where the most important ones are the equality
of the reaction rates and the conservation of the number of particles in each reaction),
we show that among the 4096 types of the interactions which have
these properties and can be modeled by a master equation and an
appropriate set of boundary conditions, there are only 28 independent
interactions which are integrable. We find all these interactions and also their
corresponding wave functions. Some of these may be new solutions of
quantum Yang--Baxter equation.
\end{abstract}

\begin{center}
 \\
\end{center}
\end{titlepage}
\newpage
\section{ Introduction}

For non--equilibrium systems in low dimensions, an understanding
can often be
gained by studying rather simple models \cite{r1}--\cite{r4}. One of the
important examples of these systems are reaction--diffusion processes on a
one--dimensional lattice, which their dynamics are fully specified by their
master equation \cite{r5}--\cite{r6}. In some cases, it is possible to solve
the master equation exactly. In recent years, there has been enormous
progress in the field of exactly solvable non--equilibrium processes. These
developments were mainly triggered by the observation that the Liouville
operator of certain (1+1)--dimensional reaction--diffusion models may be
related to Hamiltonians of previously known quantum spin systems \cite{r7},
\cite{r8}.

One of the simplest examples of reaction--diffusion processes are Asymmetric
Simple Exclusion Processes (ASEP) \cite{r2}, \cite{r9}, \cite{r10}, which
has been used to describe various problems in different fields of interest,
such as the kinetics of bipolymerization\cite{r11}, dynamical models of
interface growth \cite{r12}, and traffic models\cite{r13}. The totally ASEP
model has been solved exactly by imposing the appropriate boundary condition
on the probabilities appear in the master equation\cite{r14}. The
totally ASEP model describes a process in which each lattice site can be
occupied by at most one particle and the particle hops with rate one to its
right neighboring site if it is not already occupied, otherwise the
attempted move is rejected.

There are some other interesting and more complicated processes which can be
solved by the method developed in \cite{r14}, namely by choosing a suitable
boundary condition for the master equation. For example, it has been shown
that the so called ``generalized totally ASEP model'' can be solved exactly
by this method \cite{r15}. In this model, even if the right neighboring
site of a particle is occupied, the particle hops to the next right site
by pushing all the neighboring particles to their next right sites, with a
rate which depends on the number of right neighboring particles. This model
has been further generalized in \cite{r16} by considering both the right and
left hopping of the particles.

In this paper we are going to consider a class of integrable models in which
there are {\it two} species of particles which can hop to their right neighboring
sites if those are not occupied, and also the particles interact with each other if
they are in adjacent sites. The details of this
nearest--neighboring interaction depends on the specific considered model
(see \cite{r18}--\cite{r20} for some recent works in two-- and three--species
reaction--diffusion processes). The
important point in integrable reaction--diffusion processes with more than
one type of particle is that, as we will show, the two--particle S--matrix
of the reaction, which specify the $N$--point functions, must satisfy the
Quantum Yang--Baxter Equation (QYBE). Therefore, as we expect, the number of
integrable models, in the sense that its $N$--particle S--matrix can be
factorized into a product of two--particle S--matrices, is very few. In this
paper we will find all two--species integrable reaction--diffusion
processes which have the following properties:\\
1. the particles hop to their right neighboring sites if these sites are not
occupied,\\
2. the interaction occurs only between nearest--neighbor particles,\\
3. the particles can be annihilated or created, with the only restriction
that the total number of particles is fixed,\\
4. all the interactions, including diffusions, occur with the same rate.\\
We show that among the $2^{12}=4096$ types of the interactions which have the
above--mentioned properties and can be modeled by a master equation and an
appropriate boundary condition, there are only 42 interactions which are
integrable (their two--particle S--matrices satisfy the QYBE), and from these
42 interactions, only 28 of them are independent. Some of these may be
new solutions of QYBE.

The plan of the paper is as following. In section 2, we introduce the first
kind of this interactions, which was our initial motivation in this work, in
which besides the usual hopping, the two types of particles interact as :
$A+B\rightarrow B+B$ and $B+A\rightarrow B+B.$ We show that this
interaction can be modeled by the usual master equation of ASEP and four
boundary conditions. We also show that the model is integrable. Note that
one can look
at this model (see eq.(\ref{1})) as a simple one--dimensional model of spread
of disease. If we consider $A$ particles to be the healthy individuals and the $B$
particles the diseased ones, then we expect that when $A$ and $B$ particle
are near to each other, healthy one may become diseased (in other words $B$
transmits disease to $A$). In section 3, we compute the exact two--particle
conditional probabilities of this interaction and study the long--time
behavior of this probabilities. And finally in section 4, we investigate the
class of integrable models which have the four above--mentioned properties
and deduce that there are 28 different models, which the totally ASEP model
and our first model introduced in section 2, are two of them.

\section{$AB\rightarrow BB$ and $BA\rightarrow BB$ reaction diffusion
process}

\subsection{The master equation}

Suppose there are $N$ particles of two types $A$ and $B$ on an infinite one
dimensional lattice, with interactions
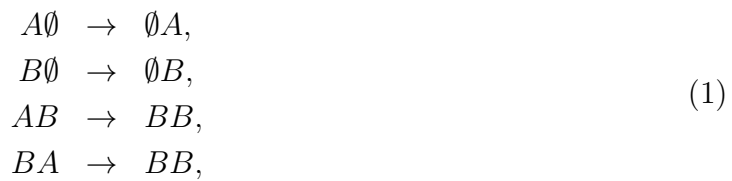
\begin{equation}
\begin{array}{rcl}
A\emptyset & \rightarrow & \emptyset A, \\
B\emptyset & \rightarrow & \emptyset B, \\
AB & \rightarrow & BB, \\
BA & \rightarrow & BB,
\end{array}
\label{1}
\end{equation}
all occur with \emph{equal} rate, which can be scaled to one. In eq.(\ref{1}), we denote the
vacancy by notation $\emptyset$. The basic quantities we are interested in are the
probabilities $P_{\alpha _1,\alpha _2,\cdot \cdot \cdot ,\alpha
_N}(x_1,x_2,\cdot \cdot \cdot ,x_N;t)$ for finding at time $t$ a particle of
type $\alpha _1$ at site $x_1$, a particle of type $\alpha _2$ at site $x_2$,
etc. Each $\alpha _i$ can be $A$ or $B$. Following \cite{r14}, we take these
functions to define probabilities only in the physical region $x_1<x_2<\cdot
\cdot \cdot <x_N$. The surfaces where any of the two adjacent coordinates
are equal, are the boundaries of the physical region. In the subset of the
physical region where $x_{i+1}-x_i>1,\forall i$, we have only hopping of the
particles and therefore the master equation can be written as
\begin{equation}
\begin{array}{rcl}
\frac \partial {\partial t}P_{\alpha _1,\alpha _2,\cdot \cdot \cdot ,\alpha
_N}(x_1,x_2,\cdot \cdot \cdot ,x_N;t) & = & P_{\alpha _1,\cdot \cdot \cdot
,\alpha _N}(x_1-1,x_2,\cdot \cdot \cdot ,x_N;t)+\cdot \cdot \cdot + \\
&  & +P_{\alpha _1,\cdot \cdot \cdot ,\alpha _N}(x_1,x_2,\cdot \cdot \cdot
,x_N-1;t) \\
&  & -NP_{\alpha _1,\cdot \cdot \cdot ,\alpha_N}(x_1,\cdot \cdot \cdot
,x_N;t).
\end{array}
\label{2}
\end{equation}
As is clear from eq.(\ref{2}), when $x_{i+1}=x_i+1$ for some $i$'s, the one
or more of the probability functions go out from the physical region and
therefore the eq.(\ref{2}) has to be supplemented by some boundary
conditions. The particular choice of the boundary condition depends on the
details of the interactions of particles. For example, it can be shown that
in the totally ASEP model, the suitable boundary condition is \cite{r14},
\begin{equation}
P(x,x)=P(x,x+1),\qquad \forall x,  \label{3}
\end{equation}
in which the time variable and also all the other coordinates have been
suppressed for simplicity. The master equation (\ref{2}) and the boundary
condition (\ref{3}) replace the very large number of equations which one
should write by considering the multitude of cases which arises in different
possible configurations.

To model the interaction (\ref{1}), we claim that the suitable boundary
conditions are :
\begin{equation}
P_{AA}(x,x)=P_{AA}(x,x+1),  \label{4-a}
\end{equation}
\begin{equation}
P_{BB}(x,x)=P_{BB}(x,x+1)+P_{AB}(x,x+1)+P_{BA}(x,x+1),  \label{4-b}
\end{equation}
\begin{equation}
P_{AB}(x,x)=P_{BA}(x,x)=0,  \label{4-c}
\end{equation}
where we have again suppressed the positions of all the other particles. By
looking at (\ref{1}), it is obvious that if we have only $A$ particles , the
process is exactly the same as totally ASEP. It is the reason of appearing
eq.(\ref{4-a}) which is the same as (\ref{3}). To justify the other three
boundary conditions, we provide a few examples in the two-- and
three--particle sectors, instead of giving a general proof.

First we consider the two--particle sector, for example $P_{BA}(x,x+1)$. By
master equation (\ref{2}) we have
\begin{equation}
\frac \partial {\partial
t}P_{BA}(x,x+1)=P_{BA}(x-1,x)+P_{BA}(x,x)-2P_{BA}(x,x+1).  \label{5}
\end{equation}
Using (\ref{4-c}), (\ref{5}) reduces to
\begin{equation}
\frac \partial {\partial t}P_{BA}(x,x+1)=P_{BA}(x-1,x)-2P_{BA}(x,x+1).
\label{6}
\end{equation}
This is exactly what we expect, as the source of configuration
$(\emptyset BA\emptyset)$ is
$(B\emptyset A\emptyset )$ and its sinks are two configurations
$(\emptyset B\emptyset A)$ and $(\emptyset BB\emptyset ).$ The
second example is $P_{BB}(x,x+1)$. Using again the master equation (\ref{2})
and the boundary condition (\ref{4-b}), we obtain
\begin{equation}
\frac \partial {\partial
t}P_{BB}(x,x+1)=P_{BB}(x-1,x)+P_{AB}(x,x+1)+P_{BA}(x,x+1)-P_{BB}(x,x+1).
\label{7}
\end{equation}
This equation also predicts the true source and sink terms, because
$(\emptyset BB\emptyset)$
has three sources $(B\emptyset B\emptyset),(\emptyset AB\emptyset),$
and $(\emptyset BA\emptyset)$ and only one sink $(\emptyset B\emptyset B).$
As a three--particle sector example, let us consider the most nontrivial case
$P_{BBB}(x,x+1,x+2).$ Using (\ref{3}) and (\ref{4-b}), we find
\begin{equation} \begin{array}{c}
\frac \partial {\partial t}P_{BBB}(x,x+1,x+2)=P_{BBB}(x-1,x+1,x+2)+P_{BBB}(x,x,x+2)
+ \\ P_{BBB}(x,x+1,x+1)-3P_{BBB}(x,x+1,x+2) \\ =P_{BBB}(x-1,x+1,x+2)
+2P_{BAB}(x,x+1,x+2)
+ \\ P_{BBA}(x,x+1,x+2)+P_{ABB}(x,x+1,x+2)-P_{BBB}(x,x+1,x+2).  \end{array}
\label{8} \end{equation} This is also the true equation, because the configuration
$(\emptyset BBB\emptyset)$ has five sources namely $(B\emptyset BB\emptyset),
2(\emptyset BAB\emptyset),(\emptyset BBA\emptyset)$,
and $(\emptyset ABB\emptyset)$ and one sink
$(\emptyset BB\emptyset B).$
The reason of appearing the factor 2 in $(\emptyset BAB\emptyset)$ is that
$(\emptyset BA\cdot \cdot
\cdot )$ can go to $(\emptyset BB\cdot \cdot \cdot )$ and also
$(\cdot \cdot \cdot AB\emptyset)$ can go to
$(\cdot \cdot \cdot BB\emptyset).$

\subsection{The Bethe ansatz solution}

We now try to solve the master equation (\ref{2}) with boundary conditions (
\ref{4-a})--(\ref{4-c}) by Bethe ansatz method. If we define $\Psi _{\alpha
_1,\cdot \cdot \cdot ,\alpha _N}(x_1,\cdot \cdot \cdot ,x_{N})\,\,$ through
\begin{equation}
P_{\alpha _1,\cdot \cdot \cdot ,\alpha _N}(x_1,\cdot \cdot \cdot
,x_N;t)=e^{-\epsilon _Nt}\Psi _{\alpha _1,\cdot \cdot \cdot ,\alpha
_N}(x_1,\cdot \cdot \cdot ,x_N),  \label{9}
\end{equation}
and substitute (\ref{9}) in master equation (\ref{2}) and boundary conditions
(\ref{4-a})--(\ref{4-c}), we find

\begin{eqnarray}
&&\Psi _{\alpha _1,\cdot \cdot \cdot ,\alpha _N}(x_1-1,x_2,\cdot \cdot \cdot
,x_N)+\cdot \cdot \cdot +\Psi _{\alpha _1,\cdot \cdot \cdot ,\alpha
_N}(x_1,x_2,\cdot \cdot \cdot ,x_N-1)  \nonumber  \label{10} \\
&=&(N-\epsilon _N)\Psi _{\alpha _1,\cdot \cdot \cdot ,\alpha
_N}(x_1,x_2,\cdot \cdot \cdot ,x_N),  \label{10}
\end{eqnarray}
and
\begin{eqnarray}
\Psi _{AA}(x,x) &=&\Psi _{AA}(x,x+1),  \nonumber  \label{11} \\
\Psi _{BB}(x,x) &=&\Psi _{BB}(x,x+1)+\Psi _{AB}(x,x+1)+\Psi _{BA}(x,x+1),
\label{11} \\
\Psi _{AB}(x,x) &=&\Psi _{BA}(x,x)=0.  \nonumber
\end{eqnarray}
Following \cite{r17}, it becomes easier if we use a compact notation as
follows: ${\mathbf{\Psi}}$ is a tensor of rank $N$ with components $\Psi
_{\alpha _1,\cdot \cdot \cdot \cdot ,\alpha _N}(x_1,\cdot \cdot \cdot ,x_N)$.
Then the boundary conditions (\ref{11}) can be written as
\begin{equation}
{\mathbf{\Psi}}(\cdot \cdot \cdot ,\xi ,\xi ,\cdot \cdot \cdot )={\mathbf{b}}
_{k,k+1}{\mathbf{\Psi }}(\cdot \cdot \cdot ,\xi ,\xi +1,\cdot \cdot \cdot ),
\label{12}
\end{equation}
where ${\mathbf{b}}_{k,k+1}$ is the embedding of ${\mathbf{b}}$ (the matrix
derived from (\ref{11})) in the location $k$ and $k+1$:
\begin{equation}
\begin{array}{ccc}
{\mathbf{b}}_{k,k+1}={\mathbf{1}}\otimes \cdot \cdot \cdot \otimes & \underbrace{\mathbf{b}}
& \otimes \cdot \cdot \cdot \otimes {\mathbf{1}}, \\
& k,k+1 &
\end{array}
\label{13}
\end{equation}
with
\begin{equation}
{\mathbf{b}}=\left(
\begin{array}{cccc}
1 & 0 & 0 & 0 \\
0 & 0 & 0 & 0 \\
0 & 0 & 0 & 0 \\
0 & 1 & 1 & 1
\end{array}
\right) .  \label{14}
\end{equation}
To solve eq.(\ref{10}), we write the coordinate Bethe ansatz for
${\mathbf{\Psi}}$ in the form:
\begin{equation}
{\mathbf{\Psi}}(x_1,...,x_N)=\sum_\sigma {\mathbf{A}}_\sigma
e^{i\sigma ({\mathbf{p}}).{\mathbf{x}}},  \label{15}
\end{equation}
where ${\mathbf{x}}$ and ${\mathbf{p}}$ denote the $N$--tuples coordinates and
momenta, respectively, the summation runs over all the elements of
permutation group, and ${\mathbf{A}}_\sigma$'s (tensors of
rank $N$) are coefficients that must be determined from boundary condition
(\ref{12}). Inserting (\ref{15}) into (\ref{10}), yields:
\begin{equation}
\sum\limits_\sigma {\mathbf{A}}_\sigma e^{i\sigma ({\mathbf{p}}).{\mathbf{x}}}
(e^{-i\sigma (p_1)}+\cdot \cdot \cdot +e^{-i\sigma (p_N)}+\epsilon _N-N)=0,
\label{16}
\end{equation}
from which one can find the eigenvalues $\epsilon _{N}$ as:
\begin{center}
\begin{equation}
\epsilon _N=\sum\limits_{k=1}^N(1-e^{-ip_k}).  \label{17}
\end{equation}
\end{center}
To find the coefficients ${\mathbf{A}}_\sigma $, we insert the wavefunction
(\ref{15}) in (\ref{12}), which yields
\begin{equation}
\sum\limits_\sigma e^{i\sum\limits_{j\neq k,k+1}\sigma (p_j)x_j+i(\sigma
(p_k)+\sigma (p_{k+1}))\xi }\left[ (1-e^{i\sigma (p_{k+1})}{\mathbf{b}}_{k,k+1})
{\mathbf{A}}_\sigma \right] =0.  \label{18}
\end{equation}
We note that the first part of the above equation is symmetric with respect
to interchange of $p_{k}$ and $p_{k+1}$, so if we symmetrize the bracket
with respect to this interchange, we obtain
\begin{equation}
(1-e_{}^{i\sigma (p_{k+1})}{\mathbf{b}}_{k,k+1}){\mathbf{A}}_\sigma
+(1-e^{i\sigma (p_k)}{\mathbf{b}}_{k,k+1}){\mathbf{A}}_{\sigma \sigma _k}=0,
\label{19}
\end{equation}
where $\sigma _k$ is an element of permutation group which only
interchange\thinspace $p_k$ and $p_{k+1},$ and $\sigma \sigma _k$ stands for
the product of two group elements $\sigma$ and
$\sigma_k$. Thus we obtain:
\begin{equation}
{\mathbf{A}}_{\sigma \sigma _k}={\mathbf{S}}_{k,k+1}(\sigma (p_k),\sigma
(p_{k+1})){\mathbf{A}}_\sigma ,  \label{20}
\end{equation}
where
\begin{equation}
\begin{array}{ccc}
{\mathbf{S}}_{k,k+1}(z_1,z_2)={\mathbf{1}}\otimes \cdot \cdot \cdot \otimes &
\underbrace{S(z_1,z_2)} & \otimes \cdot \cdot \cdot \otimes {\mathbf{1}}, \\
& k,k+1 &
\end{array}
\label{21}
\end{equation}
and
\begin{equation}
S(z_1,z_2)=-({\mathbf{1}}-z_1{\mathbf{b}})^{-1}
({\mathbf{1}}-z_2{\mathbf{b}}).  \label{22}
\end{equation}
In the above equations, we have denoted $e^{ip_k}$ by $z_{k}$. In
this way, we can calculate all the coefficients ${\mathbf{A}}_\sigma
 $'s in term of ${\mathbf{A}}_1$ by using eq.(\ref{20}), and \emph{it seems
that} the problem is solved for arbitrary boundary condition ${\mathbf{b}}$
matrix. But this is not the case and we should moreover check the consistency
of the solutions, which is highly nontrivial and depends on the details of
the interaction, i.e. the elements of the ${\mathbf{b}}$ matrix \cite{r17}.

To see this, let us find two coefficients ${\mathbf{A}}_{\sigma _1\sigma
_2\sigma _1}$ and ${\mathbf{A}}_{\sigma _2\sigma _1\sigma _2}.$ Note that
$\sigma _1\sigma _2\sigma _1$and $\sigma _2\sigma _1\sigma _2$ are equal as
the elements of permutation group, that is both of them when act on
$(p_1,p_2,p_3,p_4,\cdot \cdot \cdot )$ will result in
\begin{equation}
(p_1,p_2,p_3,p_4,...)\rightarrow (p_3,p_2,p_1,p_4,...).  \label{23}
\end{equation}
So we should have
\begin{equation}
{\mathbf{A}}_{\sigma _1\sigma _2\sigma _1}={\mathbf{A}}_{\sigma _2\sigma
_1\sigma _2}.  \label{24}
\end{equation}
Now using (\ref{20}), we have:
\begin{equation}
\begin{array}{rcl}
{\mathbf{A}}_{\sigma _1\sigma _2\sigma _3} & = & {S_{12}(\sigma _1\sigma
_2(p_1),\sigma _1\sigma _2(p_2)){\mathbf{A}}_{\sigma _1\sigma
_2}=S_{12}(p_2,p_3){\mathbf{A}}}_{\sigma _1\sigma _2} \\
& = & {S_{12}(p_2,p_3)S_{23}(\sigma _1(p_2),\sigma _1(p_3))
{\mathbf{A}}_{\sigma _1}=S_{12}(p_2,p_3)S_{23}(p_1,p_3)){\mathbf{A}}_{\sigma _1}} \\
& = & {S_{12}(p_2,p_3)S_{23}(p_1,p_3)S_{12}(p_1,p_2){\mathbf{A}}_1,}
\end{array}
\label{25}
\end{equation}
and in the same way ,
\begin{equation}
{\mathbf{A}}_{\sigma _2\sigma _1\sigma
_2}=S_{23}(p_1,p_2)S_{12}(p_1,p_3)S_{23}(p_2,p_3){\mathbf{A}}_1.  \label{26}
\end{equation}
Therefore, (\ref{24}) yields:
\begin{equation}
S_{12}(p_2,p_3)S_{23}(p_1,p_3)S_{12}(p_1,p_2)=S_{23}(p_1,p_2)S_{12}(p_1,p_3)
S_{23}(p_2,p_3),
\label{27}
\end{equation}
which is the familiar Quantum Yang--Baxter equation. Therefore , we must
check if the \textbf{S}--matrices defined in (\ref{21}) and (\ref{22}) , with
${\mathbf{b}}$ from eq.(\ref{14}) , satisfy (\ref{27}) or not.

Using (\ref{14}) , it can be shown that (\ref{22}) is equal to
\begin{equation}
S(z,w)=\frac 1{z-1}\left[
\begin{array}{cccc}
1-w & 0 & 0 & 0 \\
0 & 1-z & 0 & 0 \\
0 & 0 & 1-z & 0 \\
0 & z-w & z-w & 1-w
\end{array}
\right] ,  \label{28}
\end{equation}
and if we define $z=e^{ip_1},w=e^{ip_2},$ and $t=e^{ip_3}$, the eq. (\ref{27})
can be written as
\begin{equation}
(S(w,t)\otimes {\mathbf{1}})({\mathbf{1}}\otimes S(z,t))(S(z,w)\otimes
{\mathbf{1}})=({\mathbf{1}}\otimes S(z,w))(S(z,t)\otimes {\mathbf{1}})
({\mathbf{1}}\otimes S(w,t)).  \label{29}
\end{equation}
Now it is not too difficult to show that the matrix (\ref{28}) really
\emph{satisfy} eq.(\ref{29}), and therefore the solutions of interactions
(\ref{1}) {\it are} wavefunctions (\ref{15}) with the coefficients that can be found
by eq.(\ref{20}).

\section{The two--particle conditional probabilities}

In this section we want to calculate the two--particle conditional
probabilities \linebreak
$P(\alpha _1,\alpha _2,x_1,x_2;t|\beta _1,\beta _2,y_1,y_2;0),$
which is the probability of finding $\alpha _{1}$ at site $x_1
$ and particle $\alpha _2$ at site $x_2$ at time $t$, if at time $t=0$ we
have the particle $\beta _1$ at site $y_1$ and particle $\beta _2$
at site $y_2$. As has been discussed in \cite{r14} and \cite{r15} , we
should take a linear combination of eigenfunctions $P(x_1,x_2)$ (from (\ref{9})
and (\ref{15})) with suitable coefficients, to find these two--particle
conditional probabilities. Therefore,
\begin{equation}
\begin{array}{l}
\left(
\begin{array}{c}
P_{AA} \\
P_{AB} \\
P_{BA} \\
P_{BB}
\end{array}
\right) ({\mathbf{x}};t|{\bf \beta},{\mathbf{y}};0)= \\
=\int f(p_1,p_2)e^{-\epsilon _2t}{\mathbf{\Psi}}(x_1,x_2)dp_1dp_2 \\
=\frac 1{(2\pi )^2}\int e^{-\epsilon _2t}e^{-i\mathbf{p.y}}\left\{ \left(
\begin{array}{c}
a \\
b \\
c \\
d
\end{array}
\right) e^{i(p_1x_1+p_2x_2)}+S_{12}(p_1,p_2)\left(
\begin{array}{c}
a \\
b \\
c \\
d
\end{array}
\right) e^{i(p_2x_1+p_1x_2)}\right\} dp_1dp_2 .
\end{array}
\label{30}
\end{equation}
In the above expansion $P_{\alpha _1\alpha _2}({\mathbf{x}};t|{\bf \beta}
,{\mathbf{y}};0)$ is $P(\alpha _1,\alpha _2,x_1,x_2;t|\beta _1,\beta
_2,y_1,y_2;0),$ $f(p_1,p_2)$ is the coefficient of expansion which in
the second equality we choose it $\frac 1{(2\pi )^2}e^{-i{\mathbf{p.y}}},$
and $\epsilon _2=2-e^{-ip_1}-e^{-ip_2}$ (see (\ref{17})). ${\mathbf{\Psi}}
(x_1,x_2)$ is the two--particle wave function where from (\ref{15}) and (\ref
{20}) we obtain
\begin{equation}
\begin{array}{rcl}
{\mathbf{\Psi}}(x_1,x_2) & = & {\mathbf{A}}_1e^{i(p_1x_1+p_2x_2)}+
{\mathbf{A}}_{\sigma _1}e^{i\sigma _1({\mathbf{p).x}}} \\
& = & {\mathbf{A}}_1e^{i(p_1x_1+p_2x_2)}+S_{12}(p_1,p_2){\mathbf{A}}_1
e^{i(p_1x_2+p_2x_1)}
\end{array}
.  \label{31}
\end{equation}
In the two--particle sector, ${\mathbf{A}}_1$ is a 4--column vector
whose components must be determined by initial conditions, and
$S_{12}(p_1,p_2)$ can be read from (\ref{28}):
\begin{equation}
S_{12}(p_1,p_2)=\left(
\begin{array}{cccc}
s^{\prime } & 0 & 0 & 0 \\
0 & -1 & 0 & 0 \\
0 & 0 & -1 & 0 \\
0 & s & s & s^{\prime }
\end{array}
\right) ,  \label{32}
\end{equation}
with
\begin{eqnarray}
s^{\prime } &=&\frac{1-e^{ip_2}}{e^{ip_1}-1},  \nonumber \\
s &=&\frac{e^{ip_1}-e^{ip_2}}{e^{ip_1}-1}.  \label{33}
\end{eqnarray}

Let us first calculate eq.(\ref{30}) irrespective of initial conditions ,
that is for arbitrary $a,b,c,d.$ Substituting (\ref{32}) in (\ref{30}), we
find:
\begin{equation}
\left(
\begin{array}{c}
P_{AA} \\
P_{AB} \\
P_{BA} \\
P_{BB}
\end{array}
\right) ({\mathbf{x}};t|{\bf \beta},{\mathbf{y}};0)=\left(
\begin{array}{c}
a(F_1(t)+F_4(t)) \\
b(F_1(t)-F_2(t)) \\
c(F_1(t)-F_2(t)) \\
d(F_1(t)+F_4(t))+(b+c)F_3(t)
\end{array}
\right) ,  \label{34}
\end{equation}
in which
\begin{eqnarray}
F_1(t) &=&\frac 1{(2\pi )^2}\int e^{-\epsilon _2t}e^{i{\mathbf{p.(x-y)}}}
dp_1dp_2,  \label{35} \\
F_2(t) &=&\frac 1{(2\pi )^2}\int e^{-\epsilon _2t}e^{i(\widetilde{{\mathbf{p}}}%
{\mathbf{.x-p.y)}}}dp_1dp_2,  \label{36} \\
F_3(t) &=&\frac 1{(2\pi )^2}\int e^{-\epsilon _2t}\frac{e^{ip_1}-e^{ip_2}}{%
e^{ip_1}-1}e^{i(\widetilde{\mathbf{p}}{\mathbf{.x-p.y)}}}dp_1dp_2,  \label{37}
\\
F_4(t) &=&\frac 1{(2\pi )^2}\int e^{-\epsilon _2t}\frac{1-e^{ip_2}}{%
e^{ip_1}-1}e^{i(\widetilde{\mathbf{p}}{\mathbf{.x-p.y)}}}dp_1dp_2,  \label{38}
\end{eqnarray}
where in the above equations we have suppressed the ${\mathbf{x}}$ and ${\mathbf{p}}$
dependence of $F_i$'s for simplicity and $\widetilde{{\mathbf{p}}}
=(p_2,p_1).$ To avoid the singularity in $s$ and $s^{\prime }$, we set
$p_1\rightarrow p_1+i\varepsilon $ , and then by some simple contour
integration we find
\begin{eqnarray}
F_1(0) &=&\delta _{x_1,y_1}\delta _{x_2y_2},  \nonumber \\
F_2(0) &=&F_3(0)=F_4(0)=0,  \label{39}
\end{eqnarray}
and at $t\neq 0,$
\begin{equation}
\begin{array}{rcl}
F_1(t) & = & e^{-2t}\frac{t^{x_1-y_1}}{(x_1-y_1)!}\frac{t^{x_2-y_2}}{%
(x_2-y_2)!}, \\
F_2(t) & = & e^{-2t}\frac{t^{x_2-y_1}}{(x_2-y_1)!}\frac{t^{x_1-y_2}}{%
(x_{_1}-y_2)!}, \\
F_3(t) & = & e^{-2t}\left\{ \frac{t^{x_1-y_2+1}}{(x_1-y_2+1)!}%
\sum\limits_{k=0}^\infty \frac{t^{x_2-y_1+k}}{(x_2-y_1+k)!}-\frac{t^{x_1-y_2}%
}{(x_1-y_2)!}\sum\limits_{k=0}^\infty \frac{t^{x_2-y_1+k+1}}{(x_2-y_1+k+1)!}%
\right\} , \\
F_4(t) & = & e^{-2t}\left\{ \frac{t^{x_1-y_2+1}}{(x_1-y_2+1)!}-\frac{%
t^{x_1-y_2}}{(x_1-y_2)!}\right\} \sum\limits_{k=0}^\infty \frac{t^{x_2-y_1+k}%
}{(x_2-y_1+k)!}.
\end{array}
\label{40}
\end{equation}
Now we can study the different initial conditions.\\
\\
{\large\bf a) Case of $\beta _1=\beta _2=A$}\\
\\
If at $t=0,$ both particles are $A$ type, then our initial condition is
\begin{center}
\begin{equation}
\left(
\begin{array}{c}
P_{AA} \\
P_{AB} \\
P_{BA} \\
P_{BB}
\end{array}
\right) ({\mathbf{x}};0|A,A,{\mathbf{y}};0)=\left(
\begin{array}{c}
\delta _{x_1,y_1}\delta _{x_2,y_2} \\
0 \\
0 \\
0
\end{array}
\right) .  \label{41}
\end{equation}
\end{center}
Using (\ref{34}) and (\ref{39}) we find
\begin{equation}
a=1,b=c=d=0,  \label{42}
\end{equation}
and therefore
\begin{equation}
P_{AA}({\mathbf{x}};t|A,A,{\mathbf{y}};0)=F_1(t)+F_4(t),  \label{43}
\end{equation}
and all other $P$'s are zero. Note that eq.(\ref{43}) is exactly the same
conditional probability that has been found in \cite{r14} for simple ASEP
model, as we expect.\\
\\
{\large\bf b) Case of $\beta _1=A,\beta _2=B$}\\
\\
In this case the only non--zero element of conditional probability, at $t=0,$
is $P_{AB}=\delta _{x_1,y_1}\delta _{x_2,y_2}.$ Therefore one finds
\begin{equation}
b=1,a=c=d=0,  \label{44}
\end{equation}
and therefore
\begin{eqnarray}
P_{AB}({\mathbf{x}};t|A,B,{\mathbf{y}};0) &=&F_1(t)-F_2(t),  \nonumber \\
P_{BB}({\mathbf{x}};t|A,B,{\mathbf{y}};0) &=&F_3(t),  \label{45}
\end{eqnarray}
and $P_{AA}=P_{BA}=0,$ which is consistent with our processes (\ref{1}). It
can be also checked that the conservation of probability holds,
\begin{equation}
\sum\limits_{x_2=y_2}^\infty \sum\limits_{x_1=y_1}^{x_2-1}(P_{AB}+P_{BB})(%
{\mathbf{x}};t|A,B,{\mathbf{y}};0)=1,  \label{46}
\end{equation}
for arbitrary $y_1,y_2$ and $t.$\\
\\
{\large\bf c) Case of $\beta _1=B,\beta _2=A$}\\
\\
In this case, the final result is :
\begin{eqnarray}
P_{BA}({\mathbf{x}};t|B,A,{\mathbf{y}};0) &=&F_1(t)-F_2(t),  \nonumber \\
P_{BB}({\mathbf{x}};t|B,A,{\mathbf{y}};0) &=&F_3(t)  \label{47}
\end{eqnarray}
and $P_{AA}=P_{AB}=0.$\\
\\
{\large\bf d) Case of $\beta _1=\beta _2=B$}\\
\\
In this case the final result is the same as the case 3.1, as we expect,
\begin{equation}
P_{BB}({\mathbf{x}};t|B,B,{\mathbf{y}};0)=F_1(t)+F_4(t),  \label{48}
\end{equation}
and $P_{AA}=P_{AB}=P_{BA}=0.$

Another interesting quantity that can be calculated, is the long time
behavior of this functions . The only nontrivial case, is the case {\bf (b)}
(or equivalently {\bf (c)}). We expect that if at $t=0$ , we have $A$
and $B$ particles, (one healthy and one diseased individuals), the healthy one
becomes diseased finally, or we have two $B$ particles finally. In other words, we
expect that (in case {\bf (b)}),
\begin{equation}
\sum\limits_{x_2=y_2}^\infty \sum\limits_{x_1=y_1}^{x_2-1}P_{AB}({\mathbf{x}}%
;t\rightarrow \infty |A,B,{\mathbf{y}};0)\mathrm{\ }\rightarrow 0,  \label{49}
\end{equation}
and
\begin{equation}
\sum\limits_{x_2=y_2}^\infty \sum\limits_{x_1=y_1}^{x_2-1}P_{BB}({\mathbf{x}}%
;t\rightarrow \infty |A,B,{\mathbf{y}};0)\rightarrow 1.  \label{50}
\end{equation}
After some calculations , one can show that
\begin{equation}
\sum\limits_{x_2=y_2}^\infty \sum\limits_{x_1=y_1}^{x_{2-1}}P_{AB}({\mathbf{x}}%
;t|A,B,{\mathbf{y}};0)=
e^{-2t}\left[ 2\sum\limits_{m=1}^{y_2-y_1-1}I_m(2t)+I_0(2t)+I_{y_2-y_1}(2t)\right] ,
\label{51}
\end{equation}
where $I_n(x)$ is the $n$-th order Bessel function of the first kind :
\begin{equation}
I_n(x)=\sum\limits_{k=0}^\infty \frac{(x/2)^{n+2k}}{k!(n+k)!}.
\label{52}
\end{equation}
Now at $x\rightarrow \infty$, we have
\begin{equation}
I_n(x)\rightarrow \frac{e^x}{\sqrt{2\pi x}},  \label{53}
\end{equation}
therefore,
\begin{equation}
\sum\limits_{x_2=y_2}^\infty \sum\limits_{x_1=y_1}^{x_2-1}P_{AB}({\mathbf{x}};%
t\rightarrow \infty |A,B,{\mathbf{y}};0)\rightarrow \frac M{\sqrt{4\pi t}},
\label{54}
\end{equation}
which goes to zero. $M$ is the number of the $I_n(2t)$'s in the left hand side
of (\ref{51}). Using (\ref{46}), we see that both limits in (\ref{49}) and (%
\ref{50}) are satisfied.

\section{The class of models}

Now we want to find the class of all two--species integrable models which have the
four properties introduced in the introduction section. If we look at the
preceding sections, we notice that all the information about the model are
abbreviated in the ${\mathbf{b}}$ matrix (\ref{14}), because this matrix comes
from the boundary conditions (\ref{4-a}) to (\ref{4-c}), and the latter
induce our interactions. We also note that the sum of each column of
${\mathbf{b}}$ matrix is one. Now we claim that each ${\mathbf{b}}$ matrix which
has the following properties:\\
1-- the non--diagonal elements are one or zero,\\
2--the sum of elements in each column is one,\\
represents a model that its interaction(s) can be induced by the master
equation (\ref{2}) plus the boundary condition(s) which can be read from
${\mathbf{b}}.$

The reason of the first requirement is that the non--diagonal elements are
the sources of our reactions, as can be seen from the example solved in
section 2.1, and if we want all the reactions to occur with equal rate one, the
pre--factors of all source terms must be one (or zero, if we don't want the
corresponding source terms). Note that we should take all the rates equal to
each other, otherwise for the reactions we are interested in (i.e. those in
which, particles change their type), the factorization (\ref{9}) will not
yield the time independent boundary condition(s), which is wrong.

The reason for the second requirement lies in the conservation of
probability. Suppose that in the first column of ${\mathbf{b}}$, for example,
the sum of the non--diagonal
elements is $m$ and the diagonal element is $n$. So we have $m$ possible
interactions each can be a sink for $AA$. Therefore if our configuration is
$(\emptyset AA\emptyset),$ we must have $m+1$ sinks, one sink for diffusion
$(\emptyset AA\emptyset)\rightarrow
(\emptyset A\emptyset A),$ and $m$ sinks for reactions:
$(\emptyset AA\emptyset)\rightarrow (\emptyset\alpha \beta \emptyset).$
Now we consider the master equation for this process:
\begin{eqnarray}
\frac \partial {\partial t}P_{AA}(x,x+1)
&=&P_{AA}(x-1,x+1)+P_{AA}(x,x)-2P_{AA}(x,x+1)  \nonumber  \label{55} \\
&=&P_{AA}(x-1,x+1)+\sum\limits_1^{m'}P_{\alpha \beta
}(x,x+1)-(2-n)P_{AA}(x,x+1).  \label{55}
\end{eqnarray}
in which we have supposed that the $m'$ elements of the first row of
${\mathbf{b}}$ (besides b$_{11}$) are different from zero and therefore the corresponding
probabilities appear in $P_{AA}(x,x)$. Now as
the number of sinks must be $m+1,$ so $2-n=m+1$ which yields $n+m=1.$
Therefore the sum of the elements of the first column of ${\mathbf{b}}$ must
be one. By the
same reasoning, it is true for other columns.

In this way we have $2^{12}=4096$ possibilities for matrices ${\mathbf{b}}$
(there are 12 non--diagonal elements, each can be one or zero), each plus
master equation (\ref{2}) can model a reaction--diffusion process. But as we
have seen in reaction (\ref{2}), these ${\mathbf{b}}$'s must be consistent
with the QYBE (\ref{29}). Therefore the domain of ${\mathbf{b}}$'s is much
smaller. So it is sufficient to check which of these ${\mathbf{b}}$'s (or more
carefully, the S--matrices that are constructed by these ${\mathbf{b}}$'s from
eq.(\ref{22})) satisfy (\ref{29}). Using a symbolic manipulator (e.g.
MAPLE), we found that there are 42 different ${\mathbf{b}}$'s that satisfy
eq.(\ref{29}). By a closer inspection of these matrices
it is observed that 14 one of them can be obtained from the others by
interchanging $A\leftrightarrow B$, so they do not represent any new
physical interactions. The 42-14=28 ${\mathbf{b}}$'s (interactions) are as follows:
\[
\begin{array}{cc}
{\mathbf{b}}_1=\left(
\begin{array}{cccc}
1 & 0 & 0 & 0 \\
0 & 1 & 0 & 0 \\
0 & 0 & 1 & 0 \\
0 & 0 & 0 & 1
\end{array}
\right) ,\rm{pure \ \ \ diffusion} & {\mathbf{b}}_2=\left(
\begin{array}{cccc}
1 & 0 & 0 & 0 \\
0 & 0 & 0 & 0 \\
0 & 1 & 1 & 0 \\
0 & 0 & 0 & 1
\end{array}
\right) ,AB\rightarrow BA
\end{array}
\]
\[
\begin{array}{cc}
{\mathbf{b}}_3=\left(
\begin{array}{cccc}
1 & 0 & 0 & 0 \\
0 & 1 & 0 & 0 \\
0 & 0 & 0 & 0 \\
0 & 0 & 1 & 1
\end{array}
\right) ,BA\rightarrow BB & {\mathbf{b}}_4=\left(
\begin{array}{cccc}
1 & 0 & 0 & 0 \\
0 & 0 & 0 & 0 \\
0 & 0 & 1 & 0 \\
0 & 1 & 0 & 1
\end{array}
\right) ,AB\rightarrow BB
\end{array}
\]
\[
\begin{array}{cc}
{\mathbf{b}}_5=\left(
\begin{array}{cccc}
1 & 0 & 0 & 0 \\
0 & 0 & 0 & 0 \\
0 & 0 & 0 & 0 \\
0 & 1 & 1 & 1
\end{array}
\right) ,\left.
\begin{array}{c}
AB \\
BA
\end{array}
\right\} \rightarrow BB & {\mathbf{b}}_6=\left(
\begin{array}{cccc}
0 & 0 & 0 & 0 \\
0 & 0 & 0 & 0 \\
0 & 1 & 1 & 0 \\
1 & 0 & 0 & 1
\end{array}
\right) ,\left.
\begin{array}{c}
AB\rightarrow BA \\
AA\rightarrow BB
\end{array}
\right.
\end{array}
\]
\[
\begin{array}{cc}
{\mathbf{b}}_7=\left(
\begin{array}{cccc}
0 & 0 & 0 & 0 \\
0 & 0 & 0 & 0 \\
1 & 0 & 1 & 0 \\
0 & 1 & 0 & 1
\end{array}
\right) ,\left.
\begin{array}{c}
AA\rightarrow BA \\
AB\rightarrow BB
\end{array}
\right.  & {\mathbf{b}}_8=\left(
\begin{array}{cccc}
0 & 0 & 0 & 0 \\
0 & 0 & 0 & 0 \\
1 & 1 & 1 & 0 \\
0 & 0 & 0 & 1
\end{array}
\right) ,\left.
\begin{array}{c}
AA \\
AB
\end{array}
\right\} \rightarrow BA
\end{array}
\]
\[
\begin{array}{cc}
{\mathbf{b}}_9=\left(
\begin{array}{cccc}
0 & 0 & 0 & 0 \\
0 & 1 & 1 & 0 \\
0 & 0 & 0 & 0 \\
1 & 0 & 0 & 1
\end{array}
\right) ,\left.
\begin{array}{c}
BA\rightarrow AB \\
AA\rightarrow BB
\end{array}
\right.  & {\mathbf{b}}_{10}=\left(
\begin{array}{cccc}
0 & 0 & 0 & 0 \\
1 & 1 & 0 & 0 \\
0 & 0 & 1 & 1 \\
0 & 0 & 0 & 0
\end{array}
\right) ,\left.
\begin{array}{c}
AA\rightarrow AB \\
BB\rightarrow BA
\end{array}
\right.
\end{array}
\]
\begin{center}
\[
\begin{array}{cc}
{\mathbf{b}}_{11}=\left(
\begin{array}{cccc}
1 & 0 & 1 & 0 \\
0 & 0 & 0 & 0 \\
0 & 0 & 0 & 0 \\
0 & 1 & 0 & 1
\end{array}
\right) ,\left.
\begin{array}{c}
BA\rightarrow AA \\
AB\rightarrow BB
\end{array}
\right.  & {\mathbf{b}}_{12}=\left(
\begin{array}{cccc}
0 & 0 & 0 & 0 \\
1 & 1 & 1 & 0 \\
0 & 0 & 0 & 0 \\
0 & 0 & 0 & 1
\end{array}
\right) ,\left.
\begin{array}{c}
AA \\
BA
\end{array}
\right\} \rightarrow AB
\end{array}
\]
\[
\begin{array}{ll}
{\mathbf{b}}_{13}=\left(
\begin{array}{cccc}
1 & 1 & 0 & 0 \\
0 & 0 & 0 & 0 \\
0 & 0 & 0 & 0 \\
0 & 0 & 1 & 1
\end{array}
\right) ,\left.
\begin{array}{c}
AB\rightarrow AA \\
BA\rightarrow BB
\end{array}
\right.  & {\mathbf{b}}_{14}=\left(
\begin{array}{cccc}
0 & 0 & 0 & 0 \\
1 & 1 & 0 & 0 \\
0 & 0 & 0 & 0 \\
0 & 0 & 1 & 1
\end{array}
\right) ,\left.
\begin{array}{c}
AA\rightarrow AB \\
BA\rightarrow BB
\end{array}
\right.
\end{array}
\]
\[
\begin{array}{ll}
{\mathbf{b}}_{15}=\left(
\begin{array}{cccc}
0 & 0 & 0 & 0 \\
0 & 1 & 0 & 1 \\
1 & 0 & 1 & 0 \\
0 & 0 & 0 & 0
\end{array}
\right) ,\left.
\begin{array}{c}
BB\rightarrow AB \\
AA\rightarrow BA
\end{array}
\right. &
{\mathbf{b}}_{16}=\left(
\begin{array}{cccc}
1 & 0 & 0 & 0 \\
0 & -1 & 0 & 0 \\
0 & 1 & 1 & 1 \\
0 & 1 & 0 & 0
\end{array}
\right) ,
\begin{array}{c}
AB\rightarrow BB \\
\left.
\begin{array}{c}
BB \\
AB
\end{array}
\right\} \rightarrow BA
\end{array}
\end{array}
\]
\[
\begin{array}{ll}
{\mathbf{b}}_{17}=\left(
\begin{array}{cccc}
0 & 0 & 0 & 0 \\
0 & 0 & 0 & 0 \\
0 & 0 & 0 & 0 \\
1 & 1 & 1 & 1
\end{array}
\right) ,\left.
\begin{array}{c}
AA \\
AB \\
BA
\end{array}
\right\} \rightarrow BB
&
{\mathbf{b}}_{18}=\left(
\begin{array}{cccc}
1 & 0 & 1 & 0 \\
0 & 0 & 0 & 0 \\
0 & 1 & 0 & 1 \\
0 & 0 & 0 & 0
\end{array}
\right) ,\left.
\begin{array}{c}
BA\rightarrow AA \\
\left.
\begin{array}{c}
AB \\
BB
\end{array}
\right\} \rightarrow BA
\end{array}
\right.
\end{array}
\]
\[
\begin{array}{ll}
{\mathbf{b}}_{19}=\left(
\begin{array}{cccc}
1 & 1 & 1 & 0 \\
0 & 0 & 0 & 1 \\
0 & 0 & 0 & 1 \\
0 & 0 & 0 & -1
\end{array}
\right) ,\left.
\begin{array}{c}
\left.
\begin{array}{c}
AB \\
BA
\end{array}
\right\} \rightarrow AA \\
BB\rightarrow \left\{
\begin{array}{c}
AB \\
BA
\end{array}
\right.
\end{array}
\right.
&
{\mathbf{b}}_{20}=\left(
\begin{array}{cccc}
0 & 0 & 1 & 0 \\
0 & 0 & 0 & 1 \\
1 & 0 & 0 & 0 \\
0 & 1 & 0 & 0
\end{array}
\right) ,\left.
\begin{array}{c}
BA\rightarrow AA \\
BB\rightarrow AB \\
AA\rightarrow BA \\
AB\rightarrow BB
\end{array}
\right.
\end{array}
\]
\[
\begin{array}{ll}
{\mathbf{b}}_{21}=\left(
\begin{array}{cccc}
0 & 1 & 0 & 0 \\
1 & 0 & 0 & 0 \\
0 & 0 & 0 & 1 \\
0 & 0 & 1 & 0
\end{array}
\right) ,\left.
\begin{array}{c}
AB\rightarrow AA \\
AA\rightarrow AB \\
BB\rightarrow BA \\
BA\rightarrow BB
\end{array}
\right.
&
{\mathbf{b}}_{22}=\left(
\begin{array}{cccc}
0 & 0 & 0 & 1 \\
0 & 0 & 1 & 0 \\
0 & 1 & 0 & 0 \\
1 & 0 & 0 & 0
\end{array}
\right) ,\left.
\begin{array}{c}
BB\rightarrow AA \\
BA\rightarrow AB \\
AB\rightarrow BA \\
AA\rightarrow BB
\end{array}
\right.
\end{array}
\]
\[
\begin{array}{ll}
{\mathbf{b}}_{23}=\left(
\begin{array}{cccc}
0 & 0 & 1 & 0 \\
1 & -1 & 1 & 0 \\
0 & 1 & -1 & 1 \\
0 & 1 & 0 & 0
\end{array}
\right) ,
\begin{array}{c}
BA\rightarrow AA \\
AB\rightarrow BB \\
\left.
\begin{array}{c}
BA \\
AA
\end{array}
\right\} \rightarrow AB \\
\left.
\begin{array}{c}
BB \\
AB
\end{array}
\right\} \rightarrow BA
\end{array}
&
{\mathbf{b}}_{24}=\left(
\begin{array}{cccc}
-1 & 1 & 0 & 1 \\
0 & 0 & 0 & 1 \\
1 & 0 & 0 & 0 \\
1 & 0 & 1 & -1
\end{array}
\right) ,
\begin{array}{c}
\left.
\begin{array}{c}
BB \\
AB
\end{array}
\right\} \rightarrow AA \\
\left.
\begin{array}{c}
AA \\
BA
\end{array}
\right\} \rightarrow BB \\
BB\rightarrow AB \\
AA\rightarrow BA
\end{array}
\end{array}
\]
\end{center}
\[
{\mathbf{b}}_{25}=\left(
\begin{array}{cccc}
-1 & 0 & 1 & 1 \\
0 & -1 & 1 & 1 \\
1 & 1 & -1 & 0 \\
1 & 1 & 0 & -1
\end{array}
\right) ,
\begin{array}{c}
\left.
\begin{array}{c}
BB \\
BA
\end{array}
\right\} \rightarrow AA \\
\left.
\begin{array}{c}
BB \\
BA
\end{array}
\right\} \rightarrow AB
\end{array}
,
\begin{array}{c}
\left.
\begin{array}{c}
AA \\
AB
\end{array}
\right\} \rightarrow BA \\
\left.
\begin{array}{c}
AA \\
AB
\end{array}
\right\} \rightarrow BB
\end{array}
\]
\[
{\mathbf{b}}_{26}=\left(
\begin{array}{cccc}
-1 & 1 & 0 & 1 \\
1 & -1 & 1 & 0 \\
0 & 1 & -1 & 1 \\
1 & 0 & 1 & -1
\end{array}
\right) ,
\begin{array}{c}
\left.
\begin{array}{c}
BB \\
AB
\end{array}
\right\} \rightarrow AA \\
\left.
\begin{array}{c}
AA \\
BA
\end{array}
\right\} \rightarrow AB
\end{array}
,
\begin{array}{c}
\left.
\begin{array}{c}
BB \\
AB
\end{array}
\right\} \rightarrow BA \\
\left.
\begin{array}{c}
AA \\
BA
\end{array}
\right\} \rightarrow BB
\end{array}
\]
\[
{\mathbf{b}}_{27}=\left(
\begin{array}{cccc}
-1 & 1 & 1 & 0 \\
1 & -1 & 0 & 1 \\
1 & 0 & -1 & 1 \\
0 & 1 & 1 & -1
\end{array}
\right) ,
\begin{array}{c}
\left.
\begin{array}{c}
BA \\
AB
\end{array}
\right\} \rightarrow AA \\
\left.
\begin{array}{c}
BB \\
AA
\end{array}
\right\} \rightarrow AB
\end{array}
,
\begin{array}{c}
\left.
\begin{array}{c}
AA \\
BB
\end{array}
\right\} \rightarrow BA \\
\left.
\begin{array}{c}
AB \\
BA
\end{array}
\right\} \rightarrow BB
\end{array}
\]
\[
{\mathbf{b}}_{28}=\left(
\begin{array}{cccc}
-2 & 1 & 1 & 1 \\
1 & -2 & 1 & 1 \\
1 & 1 & -2 & 1 \\
1 & 1 & 1 & -2
\end{array}
\right) ,
\begin{array}{c}
\left.
\begin{array}{c}
AB \\
BB \\
BA
\end{array}
\right\} \rightarrow AA \\
\left.
\begin{array}{c}
AA \\
BB \\
BA
\end{array}
\right\} \rightarrow AB
\end{array}
,
\begin{array}{c}
\left.
\begin{array}{c}
BB \\
AA \\
AB
\end{array}
\right\} \rightarrow BA \\
\left.
\begin{array}{c}
BA \\
AA \\
AB
\end{array}
\right\} \rightarrow BB
\end{array}
\]
It should be mentioned that in above, the reaction processes of each
${\mathbf{b}}$ have been given only and the diffusion processes (which exist in
all cases)\ have been suppressed. Also note that ${\mathbf{b}}_1$ is the pure
diffusion process of \cite{r14}, and ${\mathbf{b}}_5$ is nothing but eq. (\ref
{14}).

In all the above cases the probabilities, ${\mathbf{\Psi}}$ and
${\mathbf{A}}_\sigma $ are given by (\ref{9}), (\ref{15}) and (\ref{20}),
respectively.
Obviously, $S(z,w)$ must be calculated from eq. (\ref{22}) for each case,
and then the calculations of section 3 can be repeated for them.

\vskip 1cm

\noindent{\bf Acknowledgement}

We would like to thank V. Karimipour for useful discussions, and R. Faraji--Dana
for helping us in computer programming. M. Alimohammadi would also like to thank
the research council of the
University of Tehran for partial financial support.

\end{document}